\documentclass[12pt]{article}
\usepackage{mathtext}
\usepackage{graphicx}
\usepackage[T2A]{fontenc}
\usepackage[utf8]{inputenc}
\usepackage{amsfonts}
\usepackage{amssymb}
\usepackage{amsmath}
\usepackage{pgfplots}
\usepackage{braket}
\usepackage{tikz}
\usepackage{float}
\renewcommand{\b}{\beta}
\newcommand{\g}{\gamma}

\newcommand{\s}{\sigma}
\newcommand{\la}{\lambda}
\renewcommand{\a}{\alpha}
\begin{document}
\title{Three principles of quantum computing}

\author{Yuri Ozhigov (1,2)\\
{\it 
Moscow State University of M.V.Lomonosov, 
} \\
{\it Faculty of computational mathematiccs and cybernetics,}
\\
{\it 2. Institute of physics and technology RAS of K.A.Valiev}
\\ 
}
\maketitle

\begin{abstract}
The point of building a quantum computer is that it allows to model living things with predictive power and gives the opportunity to control life. Its scaling means not just the improvement of the instrument part, but also, mainly, mathematical and software tools, and our understanding of the QC problem. The first principle of quantum modeling is the reduction of reality to finite-dimensional models similar to QED in optical cavities. The second principle is a strict limitation of the so-called Feynman principle, the number of qubits in the standard formulation of the QC. This means treating decoherence exclusively as a limitation of the memory of a classical modeling computer, and introducing corresponding progressive restrictions on the working area of the Hilbert space of quantum states as the model expands. The third principle is similarity in processes of different nature. The quantum nature of reality is manifested in this principle; its nature is quantum nonlocality, which is the main property that ensures the prospects of quantum physical devices and their radical advantage over classical ones.
\end{abstract}

\section{Introduction}

The quantum computer project (QC) from the original idea of R. Feynman (\cite{Fey},\cite{Deu}) has undergone a very significant metamorphosis, mainly due to serious efforts in the field of experiments with gates, as well as deepening our understanding of how quantum theory works in the field of complex processes.

At the end of the last century, the idea was generally accepted about the possibility of unlimited scaling of QCs by increasing the number of qubits, and about the possibility of performing unitary operations on their states according to the rules of quantum theory, so that decoherence (deviation of real evolution from the unitary law) was perceived as an annoying technical obstacle that can be overcome by error correction codes (\cite{Sh}). The basis of this interpretation of decoherence was the quantum master equation

\begin{equation}
\label{master_equation}  
i\hbar\dot{\rho}=[H,\rho]+i{\cal L}(\rho),\ \ {\cal L}(\rho)=\sum\limits_{i=1}^{N-1}\g_i(A\rho A^+-\frac{1}{2}\{ A^+A\rho,\rho A^+A\}),
\end{equation}
on the density matrix $\rho(t)$ of the system under consideration, in which the coherence violation is represented as the influence of the environment, and this influence enters into it in the form of decoherence factors $A_i$ forming an orthonormal basis of the space of Liouville operators, and their specific form does not depend on the Hamiltonian $H$ (\cite{BP}.

However, experiments of the last 20 years have clearly shown the fallacy of such a representation for quantum computing. The equation  \eqref{master_equation}, as well as its basis - the Schrodinger equation $i\hbar\dot{\Psi}=H\Psi$, is suitable only for systems whose complexity is low. They were the object of research by physicists of the 20th century, and quantum theory was tested on them and showed excellent agreement with experiments.

Applying this theory to a quantum computer takes us beyond its applicability as such. The fact is that the states of multi-qubit systems, which, according to traditional quantum theory, should arise during certain types of quantum computing, are not simple, and therefore go beyond the applicability of this theory. We are talking about so-called fast quantum algorithms, such as Grover's GSA algorithm (\cite{Gr}). These are exactly the processes predicted by the standard quantum theory, which in principle would be impossible to simulate on a classical computer if they exist in reality.

In light of this, decoherence is a sign that the mathematical constructions of quantum mechanics go beyond the scope of their applicability. And these limits are determined by the capabilities of classical computers. Therefore, the definition of complexity should proceed from these possibilities. Such a turn is unusual for physics, where the concept of complexity has always been secondary compared to general laws.

We will formulate a new approach to quantum theory suitable for its application to complex systems, introducing gradual restrictions on the mathematical formalism applied in it. These restrictions are already actually applied in real computations, but it is necessary to consider them as a new formalism, and not an addition to the standard one, because the field of complex systems and processes dictates its own laws.

\section{Finiteness of quantum theory}

Quantum mechanics does not accept infinities, and therefore is not fully compatible with mathematical analysis. This is well known: the eigenstates of the coordinate and momentum operators are not normalizable, and therefore the idea of an infinite-dimensional Hilbert space of quantum states contradicts the Born rule - the cornerstone principle of quantum theory. Ignoring mathematical correctness, traditionally allowed by physicists, did not bring much inconvenience in the study of simple systems and has always been regarded as a kind of curiosity. However, this is no longer acceptable for complex systems. When it comes to the limits of the applicability of formalism, we must strictly follow the requirement of correctness.

So, any state space of a quantum system must be finite-dimensional. This immediately makes it impossible to accurately move to the limit of $dx\rightarrow 0$ and lowers analytical methods to approximate ones. This technique corresponds to quantum electrodynamics (QED), in which the divergence of the series for the amplitudes of fundamental processes is solved by the so-called renormalization of charges and masses, the correctness of which is proved in \cite{BoP}. The charges and masses of elementary particles turn out to depend on the choice of a grain of resolution $dx$, which gives this grain a physical meaning.

If there is an abstraction - a one-dimensional wave function $\Psi(x)$ from a continuous variable $x$, it can be made realistic if we introduce a discrete set of possible values of the variable $x=x_0,x_1=x_0+dx,x_2=x_0+2dx,...,x_{N}=x_0+Ndx$, and then represent approximately this continuous function $\Psi(x)$ as
\begin{equation}
\Psi(x)\approx \sum\limits_{j=0}^{N-1}\Psi(x_j)d_j(x),
\label{discr}
\end{equation}
where $d_j(x)$ is the characteristic function of the $j$-th segment $[x_j,x_{j+1}],\j=0,1,...,N-1$ (see figure \ref{fig:discr1} upper part), the orthonormal basis of the $N$-dimensional the state space will consist of vectors $|j\rangle=d_j/\sqrt{dx}$, and $\la_j=\Psi(x_j)\sqrt{dx}$.

So we will come to the representation of our function in the form of a finite-dimensional state vector 
\begin{equation}
\label{state_vector}
|\Psi\rangle=\sum\limits_j\la_j|j\rangle;
\end{equation}
now this vector will be the most accurate representation of the real state for complex systems, so that the continuous function $|\Psi(t)$ will already be an approximation. (For a wave function defined on the space $R^2$ or $R^3$, instead of $\sqrt{dx}$, there will be $\sqrt{dx^2}$ or $\sqrt{dx^3}$, respectively)\footnote{The transition from discrete to continuous recording consists in the fact that all sums are replaced by integrals, and summation variables are replaced by integration variables. For example, the formula \eqref{discr} will turn into $\Psi(x)=\int\limits_R \Psi(y)\delta_y(x)dy$ where $\delta_y(x)$ is the limit of functions $d_j(x)$ at $dx\rightarrow 0$, so $x_j\rightarrow y$. Such a limit, of course, does not exist in mathematical analysis - among ordinary functions, since at $dx\rightarrow 0$ the function $d_j(x)$ will turn into a needle infinitely high and infinitely thin. This is a generalized Dirac function.}.

The discrete form of the Fourier transform and its inverse are operators acting on the basis states of an $n$-qubit system as follows:
\begin{equation}
\begin{array}{ll}
&QFT: |c\rangle\rightarrow \frac{1}{\sqrt N}\sum\limits_{a=0}^{N-1}exp(-2\pi i ac/N)|a\rangle\\
&QFT^{-1}: |a\rangle\rightarrow \frac{1}{\sqrt N}\sum\limits_{c=0}^{N-1}exp(2\pi i ac/N)|c\rangle\end{array}
\label{Fourier}
\end{equation}

Both of these mutually inverse operators with linear extension to the entire space of quantum states $C^N$ will give unitary operators - Fourier and inverse to it.

For applications, it is convenient to assume that for the variable $a$, the number $a/\sqrt{N}$ is the coordinate belonging to the segment $[0,\sqrt{N}]$ (Planck's constant in the proper system of units can be considered as a unit). Then $c/\sqrt{N}$ must be associated with the momentum. It is natural to assume that the momentum belongs to the segment $[-\sqrt{N}/2,\sqrt{N}/2]$, since a particle located on the segment $[0,\sqrt{N}]$ can move in both directions. Therefore, the momentum should be equal to $\sqrt{N}(c/N-1/2)$.

Accordingly, the discrete form of the momentum operator will be the $N$-dimensional Hermitian operator \newline $p_{discr}=QFT^{-1}\sqrt{N}(x_{discr}-I/2)QFT=A^{-1}QFT^{-1}\sqrt{N}x_{discr}QFT\ A$, where the diagonal operator $A=diag(exp(\pi i a))_{a=0,1,...,N-1}$. Its eigenvectors will have the form $A^{-1}QFT^{-1}|a\rangle$ and their eigenvalues will be the numbers $\sqrt{N}(a-1/2);\a=0,1/N,...,(N-1)/N$.

Such a discrete representation was used in \cite{Za},\cite{Wie} to construct a quantum algorithm that simulates the unitary evolution of a multiparticle system with memory growing linearly from the size of the system, and quadratic deceleration compared to real time.

So, in the discrete representation, all the eigenstates of the basic operators are normalized by one, and there are no contradictions with mathematical analysis. Here we used the analytical technique of continuous Fourier transforms to correctly write its discrete analog. It is not difficult to show that all the useful properties of the Fourier transform: the transition from differentiation (application of the momentum operator) to multiplication by a constant, as well as the identification of the hidden period of the complex exponent will be preserved during the transition from a continuous form to a discrete one, so that we can use discrete operators in finite-dimensional spaces in all physical problems related to the quantum theory of complex systems.

Note that this limitation of the formalism does not yet affect the amplitudes: they can be arbitrarily small, so for now mathematical analysis is applicable.

\section{Reduction of complexity by canonical transformation}

The second stage of formalism restriction will already affect the amplitudes. They cannot be infinitesimal because of the Born rule: we cannot introduce fully virtual events into the formalism, which in no way can be made explicitly observable in an experiment.

The physical experiment consists in choosing not the basis of the Hilbert space of quantum states, but the order of the basic vectors $|j\rangle$ in the expression \eqref{state_vector}. Each basic vector $|j\rangle$ is encoded by the binary string $a_0,a_1,...,a_{n-1}$ of the signs of the expansion of the approximation with an accuracy of $2^{-n}$ of the physical quantity $j_{phys}=L(j)$, where $j=\sum\limits_{i=0}^{n-1}a_i2^i$, $L$ is some linear transformation. For example, $j_{phys}$ can be the coordinate of a particle or its momentum. The order of the basic vectors $|j\rangle$ is always determined by the lexicographic ordering of strings $\bar a$. Thus, changing the order of the basic vectors means replacing the bits of $a_j$ with a new system of bits of $b_j$.

For example, for a system of interacting harmonic oscillators with the usual coordinates $q_l$, we can do the Fourier transform over them, considering $l$ as an argument and $q_l$ as the value of the original function, so that the new coordinates obtained by the rule
\begin{equation}
\label{fou}
Q_k=A\sum\limits exp(-\alpha i\ lk)q_l,\ k=0,1,...,N-1, N=2^n,
\end{equation}
then, from the quantum theory viewpoint, this will be a permutation of the basic vectors in the Hilbert state space, since $q_l$ and $Q_k$ are basic vectors, and the expressions \eqref{fou} are a point-to-point mapping, or a basic vector to the basic vector (\cite{Ma}). Such a transformation, extended to the entire space ${\cal H}$ of the quantum states of the oscillator system, turns out to be untangling: phonons having coordinates $Q_k$ do not get entangled when quantizing the Newtonian dynamics of the oscillator system. This type of transformation is called canonical.

Naive definition of complexity $\nu(\Psi)$ of the quantum state $|\Psi\rangle$ as the maximum number of qubits in the entangled component of the tensor decomposition $|\Psi\rangle=|\psi_1\rangle|\psi_2\rangle...|\psi_s\rangle$, should be modified taking into account the possibility of a canonical transformation reducing naive complexity. Namely,
the true complexity of the $|\Psi\rangle$ state is
\begin{equation}
\label{compl}
C(\Psi)=min_{\tau\in S_N }\nu(\tau\Psi)
\end{equation}

- minimal naive complexity over all possible permutations $\tau$ of  basic vectors.

Similarly, the complexity of the Hamiltonian $H$ is determined. Its naive complexity $\nu(H)$ is the maximum number of qubits in subsystems $S_1,S_2,...,S_s$, which form a partition of the entire system of qubits with a maximum $s$, such that decomposition takes place 
\begin{equation}
\label{ham}
H=H_1\otimes I_1+H_2\otimes I_2+...+H_s\otimes I_s
\end{equation}
where $H_i$ acts on the set of qubits $S_i,\i=1,2,...,s$. Then the true complexity of the Hamiltonian $H$ is defined as $C(H)=min_{\tau\in S_N}\nu(\tau^{-1}H\tau)$. The canonical transformation $\tau$, reduces the naive complexity of the Hamiltonian to a minimum value.

It follows from the definition that if we start from the basic state $|\Psi(0)\rangle$, then in the quantum evolution induced by the Hamiltonian $H$, only states of complexity not exceeding $C(H)$ can appear. Indeed, given \eqref{ham}, we have

$$
\exp(-itH)|\Psi(0)\rangle=\tau(H_1\otimes I_1+H_2\otimes I_2+...+H_s\otimes I_s)\tau^{-1}|\Psi(0)\rangle, 
$$
and considering that $\tau^{-1}$ is an inverse permutation of the basic, we get the required. 

The simplest example of a canonical transformation into 2 qubits is the operator $CNOT$, which reduces the Hamiltonian $H$ with 4 basic states passing into each other in pairs:
$$
H=
\begin{pmatrix}
&0&1&0&1\\
&1&0&1&0\\
&0&1&0&1\\
&1&0&1&0
\end{pmatrix}
=CNOT
(\s_x^{(1)}\otimes I_2+I_1\s_x^{(2)})
CNOT.
$$ . 

The problems in which quantum physics made progress in the 20th century had - in our definition - a small complexity. Along with the mentioned system of interacting harmonic oscillators, which serves as the basis of solid state physics, the transition to the impulse-energy representation of the electromagnetic field, which allows mathematically accurately describe the concept of photon, the theory of superfluidity, etc. advanced models are based precisely on the canonical transformation, which radically reduces the complexity of the problem.

Quantum computing has a very special nature. Quantum states arising during the implementation of Grover 's algorithm have the form 

\begin{equation}
\label{GSAstate}
|\Psi_{GSA}(t)\rangle= \a\sum\limits_{j\neq j_0,0\leq j<N}|j\rangle+\b|j\rangle,
\end{equation}
where $\a=cos(t)/\sqrt{N-1},\\b=sin(t)$ for some $t$, and $N=2^n$ has the maximal complexity
 $n$. 

This superposition has the property that all of its basic components, with the exception of exactly one, have the same nonzero amplitude, and this one component has a different amplitude. 

This property is preserved for any permutation of the basic vectors, that is, for any quasi-partial representation. But if this state was reducible, it would have the form

 $\la_1|i_1\rangle+\la_2|j_2\rangle+...)\otimes(\la_3|j_3\rangle+\la_4|j_4\rangle+...)$ for some basic $|j_i\rangle$, and such a superposition cannot contain exactly 2 values of amplitudes for all basic states, because there must either be at least 3 different non-zero values of amplitudes, or it must contain exactly two different non-zero amplitudes corresponding to two groups of basic vectors containing an equal number of elements. Both of these possibilities are excluded for states of the form \eqref{GSAstate}.

The complexity of this state is thus equal to $n$ if only $t\neq k\pi/2$ for any integer $k$. This state has the maximum possible complexity of all $n$ qubit states, and therefore can be used as a complexity meter.

The complexity of the quantum state is the amount of memory of the quantum processor, which is necessary to represent $|\Psi\rangle$ with the possibility of classical parallelization. The complexity of the Hamiltonian is the required amount of memory of a quantum processor designed to simulate the corresponding evolution, with the possibility of classical parallelization of the computation. 
The complexity of this state is thus equal to $n$ if only $t\neq k\pi/2$ for any integer $k$. This state has the maximum possible complexity of all $n$ qubit states, and therefore can be used as a complexity meter.

A complex state cannot be described with the same precision as a simple one. 
If the accuracy is $A(\Psi)$ of quantum description of the state $|\Psi\rangle$ we treat as the number of independent instances of systems that are in such a state and are available for measurement, we get a natural constraint of the form
\begin{equation}
A(\Psi)C(\Psi)\leq Q
\label{uncer}
\end{equation}
simply because there is a physical limit to the quantum memory available to us. Experiments suggest that the constant $Q$ does not exceed several tens. This constant can be found when implementing Grover's algorithm as the maximum number of qubits for which this algorithm is able to work.

\begin{figure}
\centering
\includegraphics[scale=0.45]{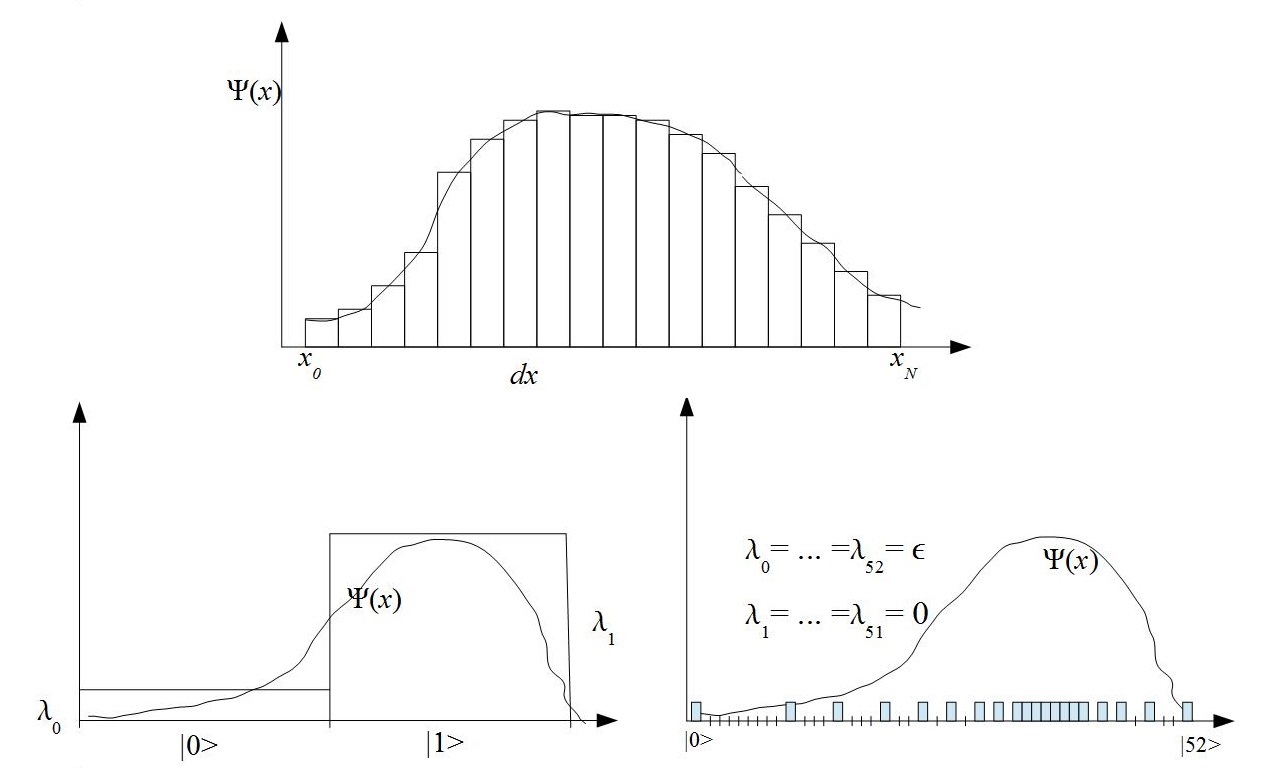}
\caption{Representation of the state vector. The curves denote the hypothetical wave function $|\Psi\rangle$ predicted by the Copenhagen theory. The rectangles represent the information about it that we can actually get.
} 
\label{fig:discr1}
\end{figure}

Given that the accuracy of determining the amplitudes of $\la_j$ in the decomposition of \eqref{state_vector} coincides with $A(\Psi)$, we get the "complexity - accuracy" dilemma, which is illustrated by the figure \ref{fig:discr1}. The case is shown at the bottom left when the main part of the computational resource is occupied with accuracy: $|\Psi\rangle=\lambda_0|0\rangle+|\lambda_1\rangle$; if we limit the number of basic states to 2, as for a particle in a two-dimensional potential, we get a satisfactory similarity with the experiment. Above is a case where the computational resource is evenly distributed between accuracy and complexity; this is the area of maximum coincidence with the experiment - a typical area of applications of quantum mechanics, where it is correct to talk about the "wave function". At the bottom right, the main resource is occupied by complexity: $|\Psi\rangle=\sum\limits_{j=0}^{N-1}\varepsilon |j\rangle,\|\varepsilon|=1$. Our knowledge here is limited to only one basic state, which is obtained in a single measurement.

Thus, the modeling of complex systems at the quantum level assumes a reduction in accuracy the greater the more complex is  the system under consideration. Already for finite-dimensional QED models (\cite{JC}) in cavities, the Hamiltonian turns out to be irreducible, and the complexity of such models grows rapidly with their size. Moreover, this will be true for the quantum description of chemistry.

Discarding states with a small amplitude is a technique that has always been used in physics. In computer modeling, this technique should be modified in such a way as to take into account the rate of amplitude growth even at very small initial values. The working area of the model, which is covered by the sum in the expansion \eqref{state_vector}, cannot be too large; it should, ideally, fit in the memory of a desktop computer.

However, there is a property of quantum computing that cannot be adequately represented using classical computing tools. This is quantum nonlocality.

\section{The principle of similarity} 

Quantum nonlocality is a subtle property of our world that gives hope for building computer models of the living. The critical quality of such models is the ultimate simplification of reality. We must discard the lion's share of the details that make up the subject of physics, chemistry and biology in order to build such a model.

The first principle of quantum computing is the reduction of models to finite-dimensional ones, which means a fundamental rejection of infinities in linear algebra, while preserving the analytical apparatus.

The second principle is the radical reduction of the dimensions of such models by means of canonical transformation, it means the rejection of mathematical analysis as a working tool; all its achievements are encapsulated in the basic states of the system. Moreover, the fact that finite-dimensional QED models are not reducible using canonical transformation suggests that all such reduction has already been done and is contained in the available finite-dimensional models. Further work will take place as a modernization of finite-dimensional models of on a computer programming platform.

But that's not enough. The enormous complexity of biological objects dictates the need for an even more radical simplification of the model. We must admit the presence of memory in the environment and evolution can no longer be considered as Markovian. We have to abandon the density matrix, moving to a virtual pure state, in which we need to include elements of the environment. Physically, this means including quantum nonlocality in the model (\cite{Aspe},\cite{Zei}). 

In chemistry and biology, complex processes are traditionally described as being controlled by classical information flows. For example, signals are interpreted as binary strings sent and received by participants of information exchange. This is a one-sided view. The control should include quantum nonlocality - only in this case the model will be predictive. Behind the complexity of the living is a physical feature that distinguishes living matter from the mechanism, and this is quantum long-range action.

The construction of such models requires reliance on the heuristics of biology, and this heuristic consists in the similarity of processes in living and inanimate matter. This similarity is provided by a single mathematical formalism - quantum mechanics in the spaces of states of strongly limited dimension. This third principle is the least developed, but it is in it that the future of quantum computing lies.

\section{Conclusion}

So, we cannot count on a big advance of the quantum computer project, betting only on the expansion of its qubit memory.
 
Three principles: finiteness, limitation of complexity by precision, and similarity form the basis for the development of a quantum operating system designed to simulate complex phenomena, primarily life, in its most complex manifestations. We should not look at a quantum computer only as a nanoelectronic device that will soon appear on our table. QC is, first of all, a method by which we hope not just to achieve a better understanding of complex processes, involving the quantum level of their description for this. It is also a way of very fine control of such processes, which makes it possible to achieve the goals in a complex way, containing unexpected and sometimes counter-intuitive moves, based, however, on a reliable foundation of quantum physics.

This is important for the proper management of the most complex system - human society. The quantum view of politics makes it objective and accessible to a wide range of participants, and avoids the costs of a head-on collision, making competition the engine of progress. This is one of the main tasks of the quantum computer project.

\section{Acknowledgements}

The work was performed at the Moscow Center for Fundamental and Applied Mathematics.

\end{document}